\documentclass[reprint,
superscriptaddress,
longbibliography,
 amsmath,amssymb,
 aps,
]{revtex4-2}

\usepackage{graphicx}% Include figure files
\usepackage{dcolumn}% Align table columns on decimal point
\usepackage{bm}% bold math
\usepackage{physics}
\usepackage{hyperref}% add hypertext capabilities
\usepackage[dvipsnames]{xcolor}
\usepackage{soul}
\usepackage{ulem}
\hypersetup{
    colorlinks=true,
    citecolor=Blue,
    linkcolor=Blue,
    urlcolor=Blue
}
\newcommand{\mref}[2]{%
    \hyperref[{#1}]{%
        \ref*{#1}#2%
    }%
}

\begin{document}
%<*document>
\title{Collisions Between Ultracold Molecules and Atoms in a Magnetic Trap}

\author{S. Jurgilas}
\author{A. Chakraborty}
\author{C. J. H. Rich}
\author{L. Caldwell}
\altaffiliation[Present Address: ]
{JILA, NIST and University of Colorado,
Boulder, Colorado 80309-0440, USA.}
\author{H. J. Williams}
\altaffiliation[Present Address: ]
{Universit\'{e} Paris-Saclay, Institut d’Optique Graduate School, CNRS, Laboratoire Charles Fabry, 91127 Palaiseau Cedex, France}
\author{N. J. Fitch}
\author{B. E. Sauer}
\affiliation{Centre for Cold Matter, Blackett Laboratory, Imperial College London, Prince Consort Road, London SW7 2AZ UK}
\author{Matthew D. Frye}
\author{Jeremy M. Hutson}
\affiliation{Joint Quantum Centre (JQC) Durham-Newcastle, Department of Chemistry, Durham University, South Road, Durham DH1 3LE, UK}
\author{M. R. Tarbutt}
\email[]{m.tarbutt@imperial.ac.uk}
\affiliation{Centre for Cold Matter, Blackett Laboratory, Imperial College London, Prince Consort Road, London SW7 2AZ UK}

\begin{abstract}
We prepare mixtures of ultracold CaF molecules and Rb atoms in a magnetic trap and study their inelastic collisions. When the atoms are prepared in the spin-stretched state and the molecules in the spin-stretched component of the first rotationally excited state, they collide inelastically with a rate coefficient $k_2 = (6.6 \pm 1.5) \times 10^{-11}\  \text{cm}^{3}/\text{s}$ at temperatures near 100~$\mu$K. We attribute this to rotation-changing collisions. When the molecules are in the ground rotational state we see no inelastic loss and set an upper bound on the spin-relaxation rate coefficient of $k_2 < 5.8 \times 10^{-12}\ \text{cm}^{3}/\text{s}$ with 95\% confidence. We compare these measurements to the results of a single-channel loss model based on quantum defect theory. The comparison suggests a short-range loss parameter close to unity for rotationally excited molecules, but below 0.04 for molecules in the rotational ground state. 
\end{abstract}

\maketitle

The formation and control of ultracold molecules is advancing rapidly, motivated by a broad range of exciting applications~\cite{Carr2009} including tests of fundamental physics, the exploration of many-body quantum physics, quantum information processing, and the study and control of chemical reactions at the quantum level. Collisions are crucial to this field, just as they have been for the field of ultracold atoms. They are a rich source of information about the interactions and reactions of atoms and molecules in a fully quantum-mechanical regime where the internal states of the reactants and the partial waves describing their relative motion are all resolved. Their control is important for evading losses and controlling reactivity~\cite{Ospelkaus2010}, and they can be harnessed for sympathetic or evaporative cooling~\cite{Son2020, Valtolina2020}. The former is especially important for laser-cooled molecules; having already produced molecules at a few $\mu$K~\cite{Cheuk2018, Caldwell2019, Ding2020}, direct laser cooling is unlikely to lower the temperature much further, and at present the densities are insufficient for evaporative cooling. Instead, sympathetic cooling with evaporatively cooled atoms~\cite{Lim2015} is a promising way to increase the phase-space density and bridge the gap to quantum degeneracy.

Several methods have been used to study collisions in the temperature range between 10~mK and 1~K. In this cold regime, crossed and merged beams have been used to study quantum state resolved collisions~\cite{Henson2012, Klein2016, Wu2017, Jankunas2015, deJongh2020} and collisions have been investigated in electric and magnetic traps loaded by buffer-gas cooling, Stark deceleration and Zeeman deceleration~\cite{Hummon2011, Parazolli2011, Fitch2020, Segev2019, Reens2017}. However, these methods are not suitable for studying collisions at $\mu$K temperatures. Molecules in this ultracold regime have been produced by atom association~\cite{Ni2008}, optoelectrical cooling~\cite{Prehn2016}, and direct laser cooling~\cite{Norrgard2016, Tarbutt2018}. For molecules produced by atom association, molecule-molecule collisions in optical traps have been an important topic of study. These collisions lead to rapid trap loss either due to chemical reactions~\cite{Ospelkaus2010} or, when reactions are energetically forbidden, due to the formation of long-lived complexes that are subsequently excited by the trapping laser~\cite{Christianen2019, Gregory2020}. The loss rate coefficients are found to be close to those predicted by a single-channel model with universal loss, in which molecules are lost with unit probability once they reach short range~\cite{Idziaszek2010, Frye2015}. Recently, reactive losses of KRb molecules have been suppressed by using an electric field and confining the molecules to two dimensions, and this has led to the formation of a stable quantum degenerate gas of these molecules~\cite{DeMarco2019, Valtolina2020, Matsuda2020}. Collisions have also been studied between ultracold CaF molecules in tweezer traps~\cite{Anderegg2019, Cheuk2020}. Rapid inelastic losses were observed in these experiments too, both for ground-state molecules and those in excited hyperfine and rotational states. Again, the loss rate coefficient was not far from the one predicted by the universal loss model.

Ultracold {\it atom-molecule} collisions have been investigated extensively by theory~\cite{Soldan2002, Krems2003, Lara2006, Wallis2011, Gonzalez-Martinez2011,Tscherbul2011}, but there are very few experimental studies. Recently, elastic collisions between optically trapped Na atoms and NaLi molecules produced by atom association were observed and used for sympathetic cooling of the molecules~\cite{Son2020}. Similarly, elastic collisions of KRb molecules with K atoms have been shown to maintain thermal equilibrium in the formation of a quantum-degenerate Fermi gas~\cite{Tobias2020}. Mixtures of laser-cooled molecules and atoms present exciting new opportunities to study and exploit ultracold collisions. Here, we produce the first such mixture and use it to study inelastic atom-molecule processes in the $\mu$K regime. We produce laser-cooled CaF molecules and Rb atoms, load them into a magnetic trap, and measure the collision-induced loss rate from the trap. When the molecules are in a rotationally excited state, the presence of Rb increases their loss rate, which we attribute to fast rotation-changing collisions. The loss rate coefficient is close to the value predicted by a single-channel universal loss model and is not suppressed when  the atoms and molecules are in spin-stretched states. By contrast, when the molecules are in the ground rotational state, no collision-induced losses are observed. We use this observation to set an upper limit to the spin-relaxation rate coefficient.  Our development of atom-molecule mixtures, and study of the inelastic processes within these mixtures, are important steps towards sympathetic cooling. 

The starting point of the experiments is a dual-species magneto-optical trap (MOT) of CaF molecules and $^{87}$Rb atoms. Each experiment begins by accumulating Rb atoms from a 2D MOT into the dual-species 3D MOT at a rate of about $2 \times 10 ^{9}$ atoms/s. Once the desired number of atoms has been loaded, the 2D MOT is turned off and the CaF MOT is loaded using the methods described previously. In brief, a pulse of CaF molecules with a mean speed of about 160~m/s is produced by a cryogenic buffer-gas source~\cite{Truppe2017c}, decelerated to low speed by frequency-chirped laser slowing~\cite{Truppe2017}, then captured into a dc MOT~\cite{Truppe2017b, Williams2017}. To lower the temperature of the molecules, we ramp down the intensity of the main CaF cooling laser to 20\% of its initial value over 4~ms, then hold it at this value for 10~ms. An image of the CaF MOT is acquired during this 10~ms period, which we use to determine the initial number of molecules, $N_{\text{CaF}}^{\rm MOT}$. Next, the quadrupole magnetic field is turned off and both species are cooled simultaneously in two independent optical molasses for 10~ms. For CaF, we follow the molasses procedure described in Ref.~\cite{Truppe2017b}. For Rb we linearly ramp the detuning and intensity of the cooling light to $-58$~MHz and 0.36~mW~cm$^{-2}$ over the 10~ms period.
The molecules cool to $100 ~\mu \text{K}$, and the atoms to $40 ~\mu \text{K}$. 

Next, we prepare the molecules in a single, selected quantum state $\ket{N,F,M_F}$ by optical pumping and microwave transfer~\cite{Williams2018} in an applied magnetic field of 230~mG along $z$. Here, $N$ is the rotational quantum number and $F$ and $M_F$ are the quantum numbers for the total angular momentum and its projection onto $z$. At the same time, the atoms are optically pumped into the state $\ket{F=2,M_{F}=2}$. Then, all the laser light is blocked using mechanical shutters and the magnetic quadrupole trap is turned on for a time $t$ at an axial field gradient of 30 G/cm. The trap depth is 1.5~mK. We measure the final number of molecules, $N_{\text{CaF}}(t)$, by recapturing them in the MOT and imaging their fluorescence for 10~ms. Molecules prepared in $N=0$ are transferred back to $N=1$ using a microwave pulse prior to recapture in the MOT. We measure the final number of atoms, $N_{\text{Rb}}(t)$, by taking an absorption image of the cloud shortly after releasing it from the magnetic trap. The collisional loss rate is determined by measuring the fraction of molecules remaining, $r_{\rm CaF}(t) = N_{\text{CaF}}(t)/N_{\text{CaF}}^{\rm MOT}$, as a function of $t$, both with and without atoms in the trap. Division by $N_{\text{CaF}}^{\rm MOT}$ makes the measurements immune to shot-to-shot fluctuations in the number of molecules in the MOT. Our simulations of the trap loading show that the sizes and positions of the clouds reach a steady state within 300~ms, which is the minimum value of $t$ we use.

\begin{figure}[t!]
    \includegraphics[width=\linewidth]{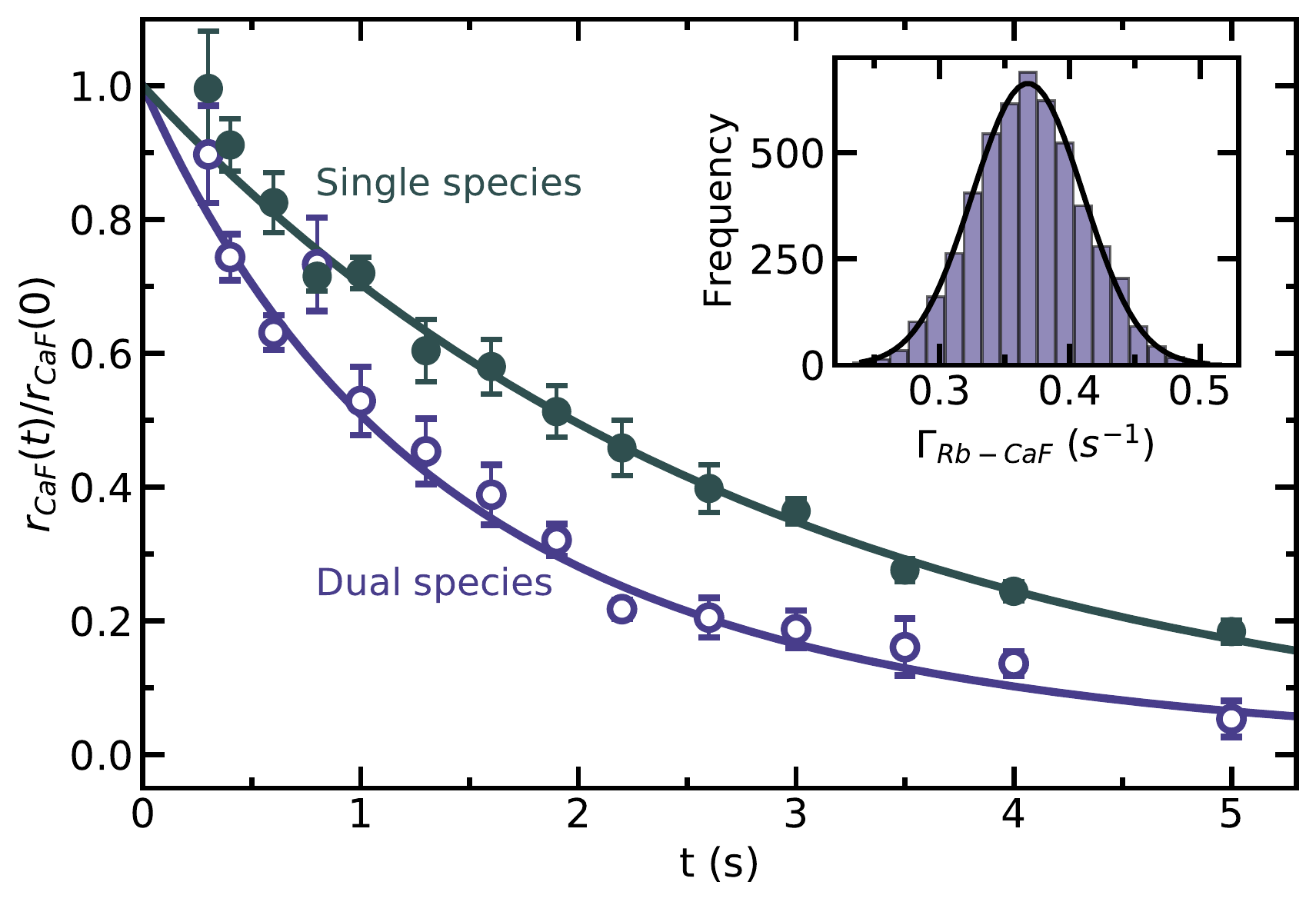}
    \caption{Fraction of molecules, $r_{\text{CaF}}(t)/r_{\text{CaF}}(0)$, remaining in the magnetic trap after a hold time $t$, with Rb (open points) and without Rb (filled points). The molecules are prepared in the $\ket{1, 2, 2}$ state. Each point is the average and standard deviation of 6 measurements. For the single species data the line is a fit to an exponential decay model. For the two-species data the line is a fit to Eq.(\ref{eq:LossModel}). Inset: histogram of $\Gamma_{\text{Rb-CaF}}$ estimates using the method described in \cite{Note1}, together with a fit to a normal distribution.}
    \label{fig:loss_N1}
\end{figure}

The density of the atomic sample exceeds that of the molecules by 6 orders of magnitude, so we can ignore molecule-molecule collisions and loss of atoms due to collisions with molecules. In this case, the loss of molecules from the trap is described by
\begin{equation}
    \dot{r}_{\text{CaF}} = \left(-\Gamma_{\text{CaF}}^{0}-\Gamma_{\text{Rb-CaF}} e^{-\Gamma_{\text{Rb}} t}\right) r_{\text{CaF}}
\end{equation}
whose solution is
\begin{equation}
    r_{\text{CaF}}(t)\!=\!r_{\text{CaF}}(0) \exp\!\left(\!-\Gamma_{\text{CaF}}^{0}t - \frac{\Gamma_{\text{Rb-CaF}}}{\Gamma_{\text{Rb}}} (1 - e^{-\Gamma_{\text{Rb}}t})\!\right).
    \label{eq:LossModel}
\end{equation}
Here $\Gamma_{\text{CaF}}^{0}$ is the loss rate of CaF in the absence of Rb, $\Gamma_{\text{Rb}}$ is the loss rate of Rb, and $\Gamma_{\text{Rb-CaF}}$ is the collision-induced loss rate of CaF at $t=0$. This is
\begin{equation}
    \Gamma_{\text{Rb-CaF}} = k_2 N_{\text{Rb}}(0)\int f_{\text{Rb}}(\vec{r})f_{\text{CaF}}(\vec{r})d^3\vec{r}
    = k_2\zeta
    \label{eq:betazeta}
\end{equation}
where $k_2$ is the inelastic rate coefficient and $f_{s}(\vec{r})$ is the density distribution of species $s$ normalised such that $\int f_{s}(\vec{r})d^3\vec{r} = 1$.
The equation defines $\zeta$, an effective Rb density that accounts for the overlap between the two distributions.

Figure~\ref{fig:loss_N1} is an example of one measurement showing the fraction of molecules remaining in the trap as a function of $t$, for molecules prepared in the state $\ket{1,2,2}$. In the absence of Rb (filled points), the data fit well to a single-exponential decay with a loss rate of $\Gamma^{0}_{\text{CaF}} = 0.35(1)$~s$^{-1}$. This is the natural loss rate due to collisions with residual background gas and vibrational excitation by blackbody radiation~\cite{Williams2018}. The Rb data, $N_{\text{Rb}}(t)$, also fit well to a single exponential with $\Gamma_{\text{Rb}} = 0.313(5)$~s$^{-1}$. These single-species loss rates vary negligibly throughout the experiments. The open points show the loss of molecules in the presence of $N_{\text{Rb}}(0)=1.50(15)\times 10^9$ atoms with a peak number density of $5(1) \times 10^{10}\ \text{cm}^{-3}$. We fit these data to Eq.~(\ref{eq:LossModel}) with $r_{\text{CaF}}(0)$ and $\Gamma_{\text{Rb-CaF}}$ as free parameters. We account for the uncertainties in $\Gamma_{0}^{\text{CaF}}$ and $\Gamma_{\text{Rb}}$ using the method described in the Supplemental Material~\footnote{\label{supp-matt-note}See Supplemental Material for details of absorption imaging, the equilibrium distribution in the magnetic trap, analysis of uncertainties, and calculation of loss rates, including Refs.~[44--50]}\nocite{Gao1998, Gao:AQDTroutines, Gao2008, Gribakin1993, Christianen(1)2019, FryeUnpublished2021, Charron1995}, which yields the  distribution of $\Gamma_{\text{Rb-CaF}}$ values shown in the inset of the figure. The mean and standard deviation of this distribution give our best estimate for this dataset, $\Gamma_{\text{Rb-CaF}} = 0.368(44)~\text{s}^{-1}$.

\begin{figure}[t!]
    \includegraphics[width=\linewidth]{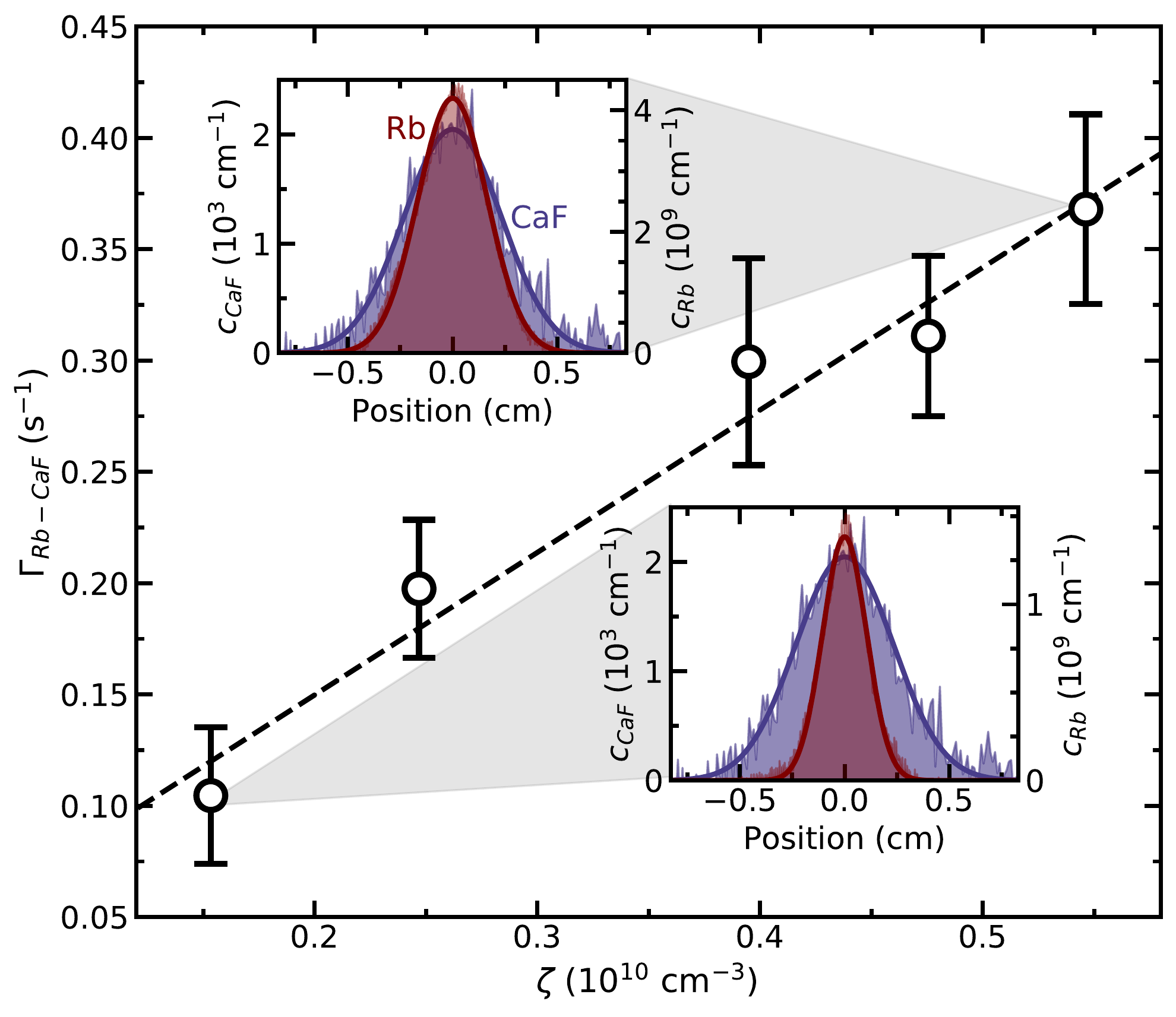}
    \caption{Points: collisional loss rate, $\Gamma_{\text{Rb-CaF}}$, as a function of effective density, $\zeta$, for molecules in the $\ket{1,2,2}$ state. Line: straight line fit to the data. Insets: radial density distributions, $c_{s}(x)=\int N_{s} f_s(x,y,z)\,dy\,dz $, of the atoms and molecules for small and large $\zeta$, with Gaussian fits.}
    \label{fig:extract_beta}
\end{figure}

To determine $\zeta$, we measure the density distributions of both species in the magnetic trap by turning off the magnetic field gradient and then imaging their distributions. For Rb we use absorption imaging as described in~\cite{Note1}. For CaF, we turn on the MOT light for 1~ms, with the field gradient still off, and image the fluorescence. Expansion of the cloud during this short period is negligible. We fit the radial distributions to Gaussians and the axial distributions to the equilibrium distribution for a magnetic quadrupole trap plus gravity~\cite{Note1}. We use these fits to calculate the overlap integral in Eq.~(\ref{eq:betazeta}). Our choice of axial distribution automatically accounts for the differential gravitational sag between the two clouds. The temperatures in the magnetic trap are measured using the standard ballistic expansion technique. We find that, when the CaF is loaded into the trap, its $1/e^2$ radius expands from $1.6(1)$ to $2.30(7)$~mm and its geometric-mean temperature rises to $175(37)$~$\mu$K. This is because the initial position of the CaF cloud is displaced from the trap center. The Rb cloud is well centered because of the greater flexibility in controlling the intensity balance of the Rb MOT light. Its geometric-mean temperature in the magnetic trap is $71(1)$~$\mu$K.

We repeat the measurements of $\Gamma_{\text{Rb-CaF}}$ and $\zeta$ for various values of $N_{\text{Rb}}(0)$. The Rb density distribution, $f_{\text{Rb}}$, depends on the number of atoms, so we measure it in every case.  Figure~\ref{fig:extract_beta} shows how $\Gamma_{\text{Rb-CaF}}$ varies with $\zeta$, together with the measured radial cloud distributions for the smallest and largest atom clouds used. The gradient of the straight-line fit gives the value of $k_2$. We account for the uncertainties in the measured number of atoms and cloud sizes using the method described in \cite{Note1}. The result is $k_2 = (6.6 \pm 1.5) \times 10^{-11}\ \text{cm}^{3}/\text{s}$ for the state $\ket{1,2,2}$. As discussed later, we attribute this loss to collisions with Rb that change the rotational state of the molecule from $N=1$ to $N=0$. This results in a loss of molecules for two reasons: (i) we detect molecules only in $N=1$; (ii) the collision releases 1~K of energy, which is far greater than the trap depth.

For the measurement described above, both the atoms and molecules are in spin-stretched states, an arrangement that often suppresses inelastic collisions~\cite{Hummon2011, Son2020}. To test whether that is the case here, we repeat the measurement using molecules in the $\ket{1,1^{+},1}$ state~\footnote{In the $N=1$ state there are two hyperfine components with $F=1$. Our notation $F=1^+$ refers to the component of highest energy.}, which is not spin stretched. We obtain $k_2 = (5.7 \pm 1.8) \times 10^{-11}\ \text{cm}^{3}/\text{s}$, showing that these rotation-changing collisions have no strong dependence on the choice of hyperfine or Zeeman level.

\begin{figure}[t!]
    \includegraphics[width=\linewidth]{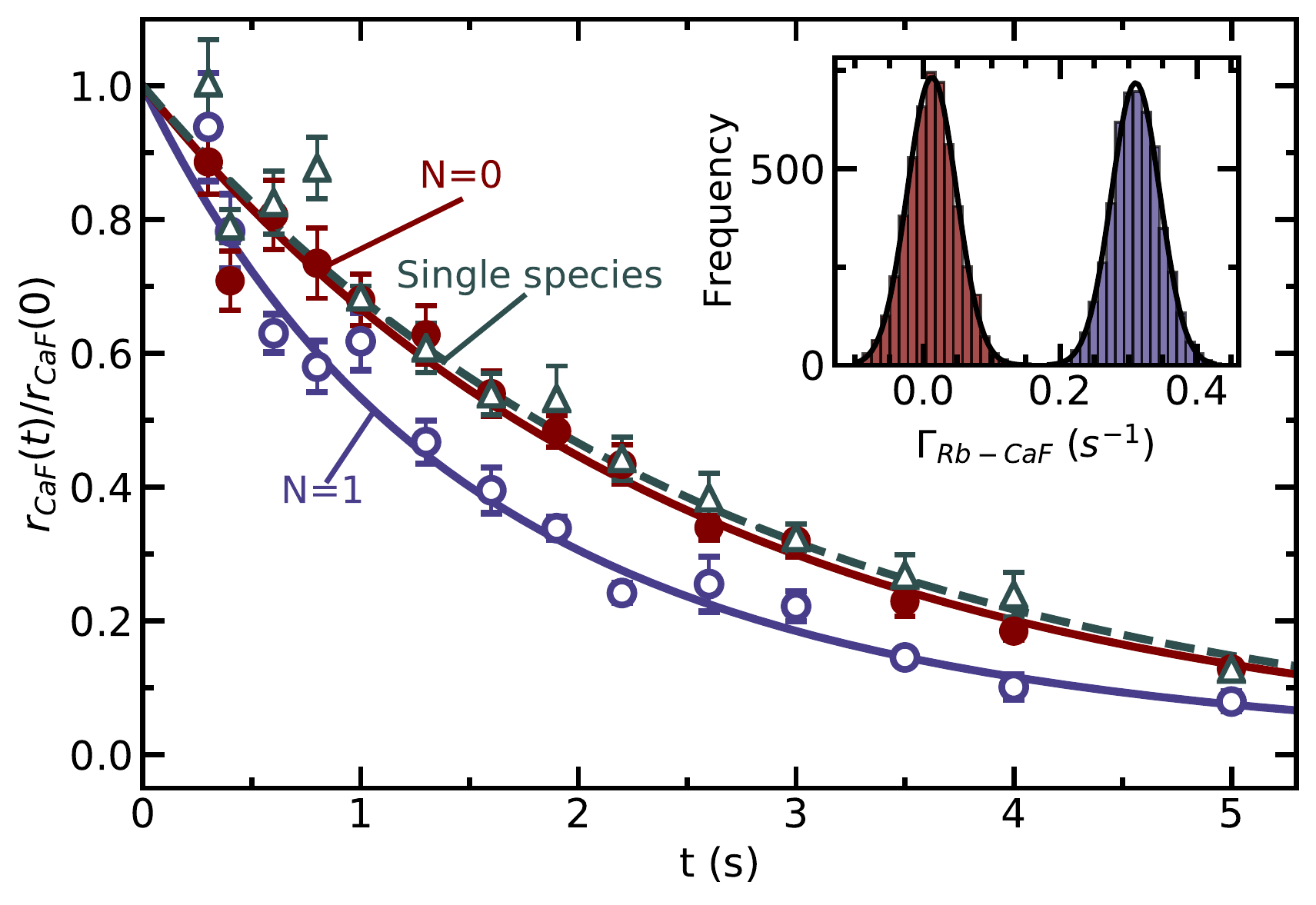}
    \caption{Loss of molecules from the magnetic trap in the presence of $1.2 \times 10^9$ atoms, for molecules in the $\ket{1,2,2}$ state (open points) and the $\ket{0,1,1}$ state (filled points). Solid lines: fits to Eq.~(\ref{eq:LossModel}). Also shown is the loss of molecules in $\ket{0,1,1}$ in the absence of Rb (open triangles) together with a single exponential fit (dashed line). Inset: histogram of $\Gamma_{\text{Rb-CaF}}$ estimates, together with fits to normal distributions.}
    \label{fig:loss_N0}
\end{figure}

Next, we make the same measurement for molecules prepared in the ground rotational state $\ket{0, 1, 1}$. Here, the only open loss channel is spin relaxation. Figure \ref{fig:loss_N0} compares the loss rates for molecules in $N=0$ and $N=1$, in the presence of $N_{\text{Rb}}(0) = 1.2 \times 10^9$ atoms, dramatically illustrating how the choice of rotational state influences these ultracold inelastic collisions. For these data, the collisional loss rate is $0.30(5)$~s$^{-1}$ for molecules in $\ket{1,2,2}$ and $0.013(36)$~s$^{-1}$ for molecules in $\ket{0,1,1}$. Averaging several similar measurements we obtain $k_2 = -0.05 \pm 2.9 \times 10^{-12}\ \text{cm}^{3}/\text{s}$. This result is consistent with zero and yields the upper bound $k_2 < 5.8 \times 10^{-12}\ \text{cm}^{3}/\text{s}$ with 95\% confidence. This is an order of magnitude smaller than the rotation-changing rate coefficient for molecules in the first excited rotational state.

\begin{figure}[tb]
    \includegraphics[width=\linewidth]{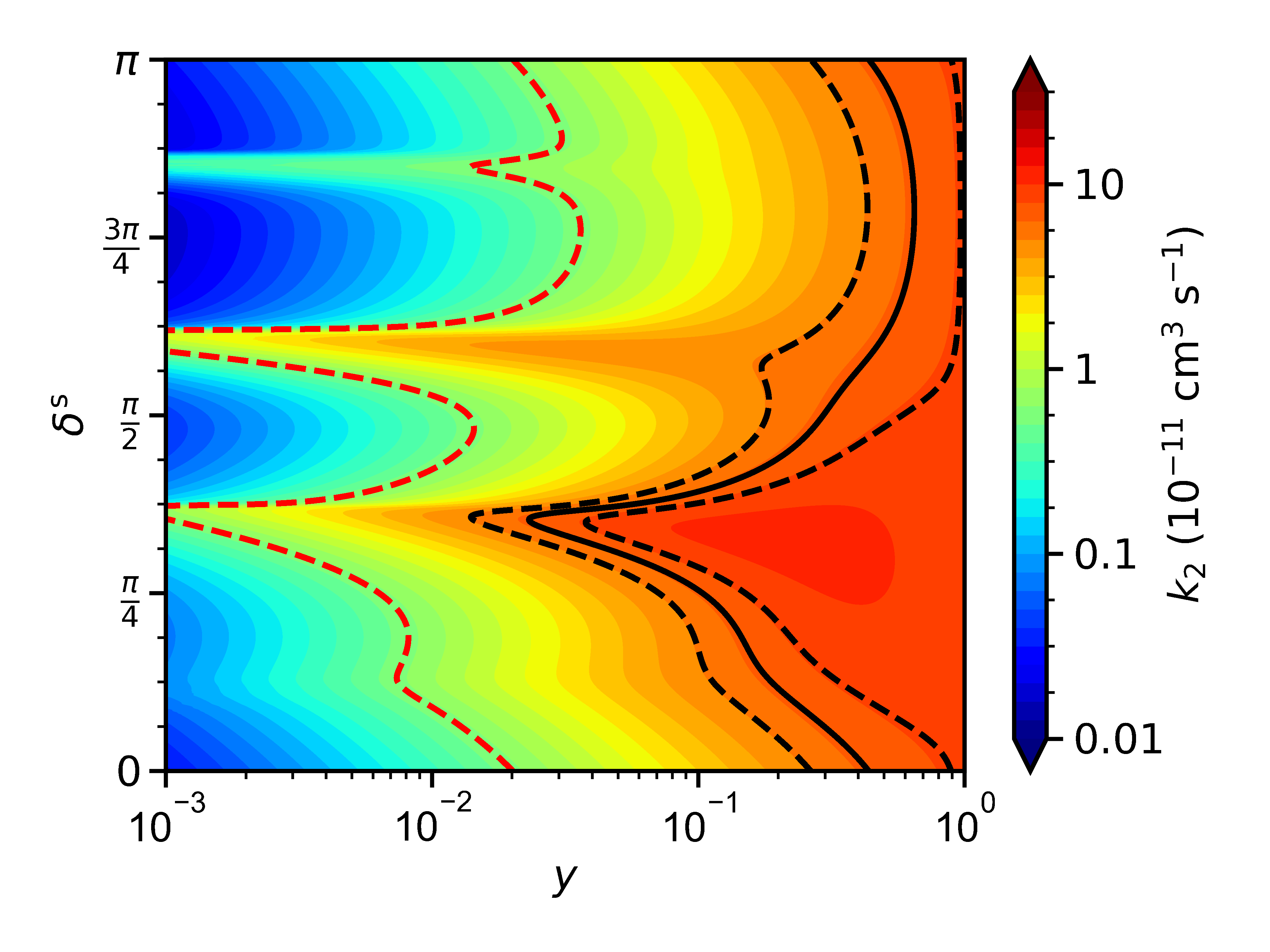}
    \caption{Thermally averaged loss rate coefficient $k_2$ from the single-channel model as a function of loss parameter $y$ and short-range phase shift $\delta^\textrm{s}$. The mean and standard error of the experimentally measured $k_2$ for molecules in $\ket{1,2,2}$ are shown as black lines. The 95\% upper bound for $k_2$ for molecules in $\ket{0,1,1}$ is shown as a dashed red line.
    \label{fig:contour}}
\end{figure}

To interpret the experimental loss rates, we compare to calculations with a single-channel non-universal model based on quantum defect theory~\cite{Frye2015,Note1}. This models the long-range interaction potential by its asymptotic form $-C_6R^{-6}$ and takes full account of temperature dependence, which is important under the conditions of the experiment. It has been used successfully to model losses in collisions of ultracold RbCs molecules~\cite{Gregory2019}.
The model characterizes the complicated short-range part of the interaction in terms of two parameters, $y$ and $\delta^\textrm{s}$. The loss parameter $y$ runs from 0 (no loss) to 1 (complete loss at short range, corresponding to a universal model). It accounts for all sources of collisional loss, including rotational inelasticity and spin-changing collisions. $\delta^\textrm{s}$ is the phase shift between the incoming wave and the reflected wave at short range, and is related to the scattering length.

Figure \ref{fig:contour} shows a contour plot of the rate coefficient obtained from the single-channel model as a function of $y$ and $\delta^\textrm{s}$. The experimental value and range of uncertainty for $\ket{1,2,2}$ and the upper limit for $\ket{0,1,1}$ are indicated as lines. The calculated loss rates show significant peaks around $\delta^\textrm{s}=3\pi/8$ and $5\pi/8$ and a weaker peak around $7\pi/8$; these are associated with p-, d-, and f-wave shape resonances, respectively. The corresponding s-wave peak around $\delta^\textrm{s}=\pi/8$, which is very prominent at lower temperatures \cite{Frye2015,Gregory2019}, is weakened at these temperatures due to increased transmission past the long-range potential and by thermal averaging.

The measured loss rate for molecules in $N=1$ is close to the universal rate, which is $8.2\times 10^{-11}$ cm$^3$ s$^{-1}$ at this temperature.
The most likely ranges of parameters have $y>0.4$, but the measurements do not rule out significantly lower $y$ if there is resonant enhancement. Since the spin state is found to have little influence on the loss rate and complex-mediated loss is unlikely to be important for this system~\cite{Note1}, we conclude that this loss is dominated by rotation-changing collisions. 
For the rotational ground state, the measured loss rate is consistent with loss parameters $y<0.04$.
Overall, the results are consistent with the expectation that rotational relaxation is very fast, but that spin relaxation is slow.  

Full coupled-channel scattering calculations including nuclear spin \cite{Gonzalez-Martinez2011} are not feasible for Rb+CaF, because the very deep potential well would require an impractically large basis set for convergence. Even without nuclear spin, they are very challenging.
Morita {\it et al.} have carried out such calculations on a single interaction potential for the related system Rb+SrF.
However, the particular interaction potential they used produces a p-wave resonance very close to threshold; it would correspond to $\delta^\textrm{s} \approx 3\pi/8$ in our model. As shown in Fig.\ \ref{fig:contour}, such a potential is likely to produce atypically large loss rates.
At a collision energy of $10^{-4}$ cm$^{-1}$ and a magnetic field of 1~G, their calculation gave a spin-relaxation cross section of $\sim200$ \AA$^2$; without thermal averaging, this corresponds to $k_2 \approx 5\times 10^{-13}$ cm$^3$ s$^{-1}$, which is well within our experimental upper bound.

In summary, we have produced the first mixtures of laser-cooled atoms and molecules and have studied their inelastic collisions in the ultracold regime. We have compared the results to a short-range loss model and find that rotational relaxation proceeds rapidly, close to the universal rate and independent of hyperfine and magnetic sub-level, whereas spin relaxation is at least 10 times slower. The latter observation is encouraging for the prospects of sympathetic cooling, where spin relaxation processes are expected to be a limiting factor.

Our mixture can be used to investigate collisions in other settings. For example, studies in the dual-species MOT can shed light on inelastic collisions with excited-state atoms and light-assisted processes~\cite{Sesko1989}. By capturing atoms and molecules from the mixture into tweezer traps, the collisions can be studied at the single-particle level~\cite{Cheuk2020}. It will be interesting to investigate the influence of applied electric and magnetic fields on these collisional processes~\cite{Parazolli2011, Gonzalez-Martinez2011}. The mixture could also be used to form ultracold triatomic molecules by photoassociation or magnetoassociation. We are currently working to trap more molecules, trap atoms at a higher density, improve the overlap of the clouds, and incorporate an optical dipole trap. With these improvements, we aim to study the elastic collisions required for sympathetic cooling of molecules by evaporatively cooled atoms~\cite{Lim2015}, which is a promising way to increase the phase-space density towards quantum degeneracy.

Underlying data may be accessed from Zenodo~\cite{zenodoCollisionsMQT2021} and used under the Creative Commons CCZero license. 

We are grateful for expert technical assistance from Jon Dyne and David Pitman. We acknowledge helpful discussions with Micha{\l} Tomza. This work was supported by EPSRC grant EP/P01058X/1.
%</document>
\bibliography{references}

\end{document}

\title{Collisions Between Ultracold Molecules and Atoms in a Magnetic Trap: \\ Supplemental Material}

\author{S. Jurgilas}
\author{A. Chakraborty}
\author{C. J. H. Rich}
\author{L. Caldwell}
\author{H. J. Williams}
\author{N. J. Fitch}
\author{B. E. Sauer}
\affiliation{Centre for Cold Matter, Blackett Laboratory, Imperial College London, Prince Consort Road, London SW7 2AZ UK}
\author{Matthew D. Frye}
\author{Jeremy M. Hutson}
\affiliation{Joint Quantum Centre (JQC) Durham-Newcastle, Department of Chemistry, Durham University, South Road, Durham DH1 3LE, UK}
\author{M. R. Tarbutt}
\affiliation{Centre for Cold Matter, Blackett Laboratory, Imperial College London, Prince Consort Road, London SW7 2AZ UK}

\maketitle
%<*document>
\section{Determining Atom density}

We determine the number densities of the trapped atom clouds by absorption imaging. A collimated laser beam traverses the atom cloud whose shadow is imaged onto a CCD camera using a pair of achromatic lenses. The first lens has a focal length of $f_{1} = 300$~mm and is a distance  $f_{1} + \delta l_{1}$ away from the atoms, while the second lens has $f_{2} = 125$~mm and is a distance $f_{2} + \delta l_{2}$ from the camera. The lenses are separated by $f_{1} + f_{2}$. Here $\delta l_{1}$ and $\delta l_{2}$ account for positioning errors in the imaging system. 
Due to its refractive index and Gaussian density distribution, the atom cloud acts like a lens whose focal length is approximately
\begin{equation}
    f_{A} = -\frac{2\pi\Gamma\sigma^2}{\Delta\lambda \tau}
\end{equation}
where $\Delta$ is the detuning of the light, $\lambda$ is the wavelength, $\Gamma$ is the decay rate of the excited state, $\sigma$ is the standard deviation of the Gaussian density distribution and $\tau$ is the optical depth at the center of the cloud. When $\Delta < 0$, $f_{A}$ is positive, meaning that the cloud acts as a converging lens.
The magnification of the entire imaging system is 
\begin{equation}
    M = \frac{\delta l_{2} f_{1}}{f_{2} f_{A}} + \frac{f_{2}(\delta l_{1} - f_{A})}{f_{1} f_{A}}.
    \label{eq:magnification}
\end{equation}

\begin{figure}[t]
    \includegraphics[width=0.9\linewidth]{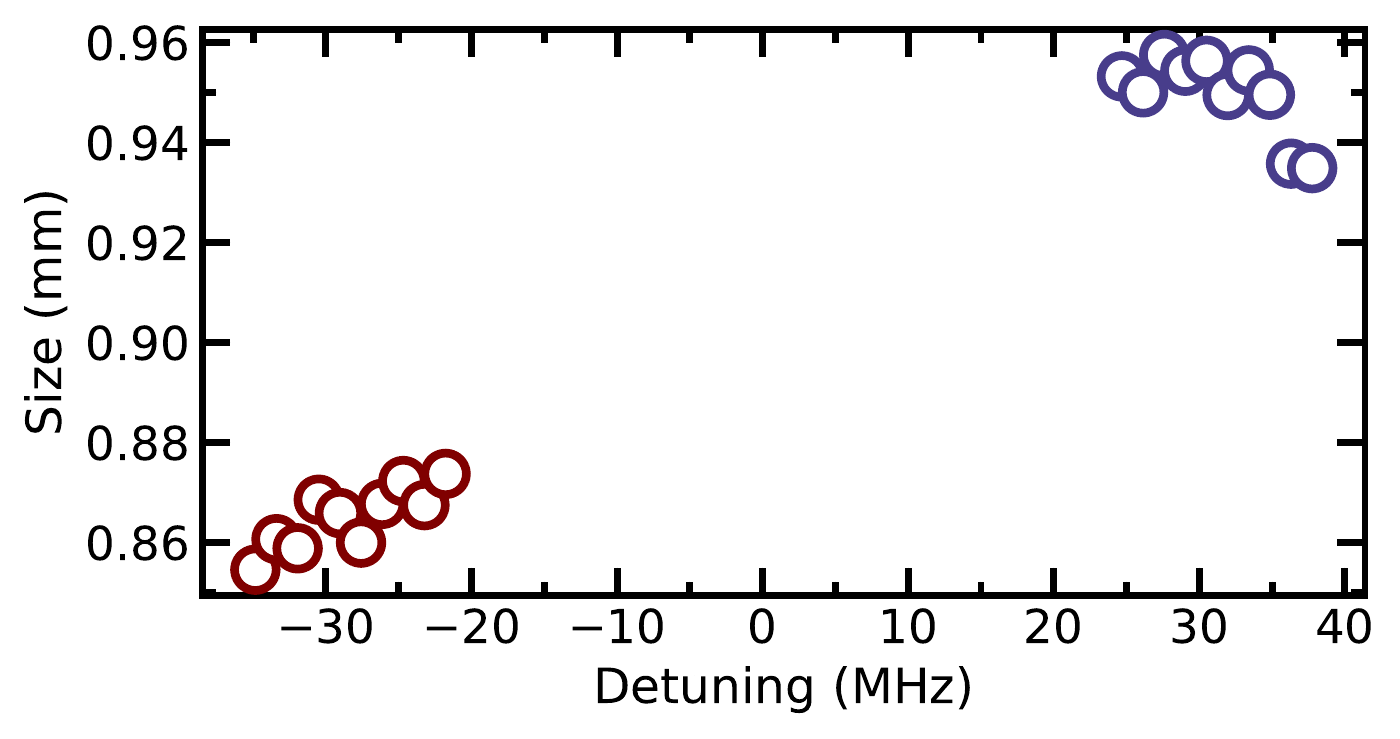}
    \caption{Apparent RMS radial size of the atom cloud for various probe detunings. The sizes are determined by fitting 2D Gaussian distributions to the acquired images.}
    \label{fig:atom_lensing}
\end{figure}

For a resonant probe, $f_{A}\rightarrow \infty$ and the magnification is simply $M=-f_{2}/f_{1}$ and is robust against $\delta l_{1}$ and $\delta l_{2}$. On resonance however, the degree of absorption by the atom cloud exceeds the dynamic range of the camera, so we take absorption images with $\Delta=-2.42\Gamma$. In this case, when all distances are set perfectly ($\delta l_{1}=\delta l_{2}=0$) the lensing effect of the atom cloud is canceled and again the magnification is $M=-f_{2}/f_{1}$. To investigate the residual systematic shift due to lensing when there are positioning errors ($\delta l_{1,2} \ne 0$), we acquire absorption images for a range of detunings. Figure~\ref{fig:atom_lensing} shows how the apparent cloud size varies with $\Delta$. As expected from the qualitative behaviour described above, there is a difference of about 10\% in apparent size between positive and negative $\Delta$ over the range explored. We conclude that in our standard absorption images the cloud size is underestimated by 3--5\% (depending on the atom density) and we apply a systematic correction to all the Rb data to account for this. This leads to a 4\% systematic correction to the values of $k_2$, which we have applied to our results. The uncertainty in making this correction contributes negligibly to the overall uncertainty of the measurement.

In the absorption imaging, the effective absorption cross section depends on the polarization of the probe beam relative to the applied magnetic field, the intensity and detuning of the probe beam, the initial state distribution of the atomic sample and the duration of the imaging pulse. We optically pump atoms to the state $\ket{F=2,M_{F}=2}$ prior to magnetic trapping, and the magnetic field gradient of the trap is slightly below the value needed to levitate atoms in $\ket{2,1}$ or $\ket{1,-1}$ against gravity. This removes any uncertainty about the state distribution. We use a set of three orthogonal Helmholtz coil pairs to cancel the background magnetic field and apply a 200~mG bias field along the direction of the probe beam. The coils are calibrated by microwave spectroscopy of the CaF molecules using a magnetically-sensitive hyperfine component of the rotational transition~\cite{Williams2018}. The probe laser beam is circularly polarized using a quarter wave plate. The variation of the optical depth with the angle of the quarter wave plate matches well with the results of a model of the absorption imaging where we solve the set of rate equations describing the atom-light interaction for various polarizations. This gives us high confidence that the imaging is well understood. In the experiments, we set the polarization to drive $\sigma^{+}$ transitions from the $\ket{2,2}$ state, so that the resonant absorption cross section is $3\lambda^2/(2\pi)$.

\section{Equilibrium distribution in magnetic quadrupole trap}

\begin{figure*}[t]
    \includegraphics[width=0.85\linewidth]{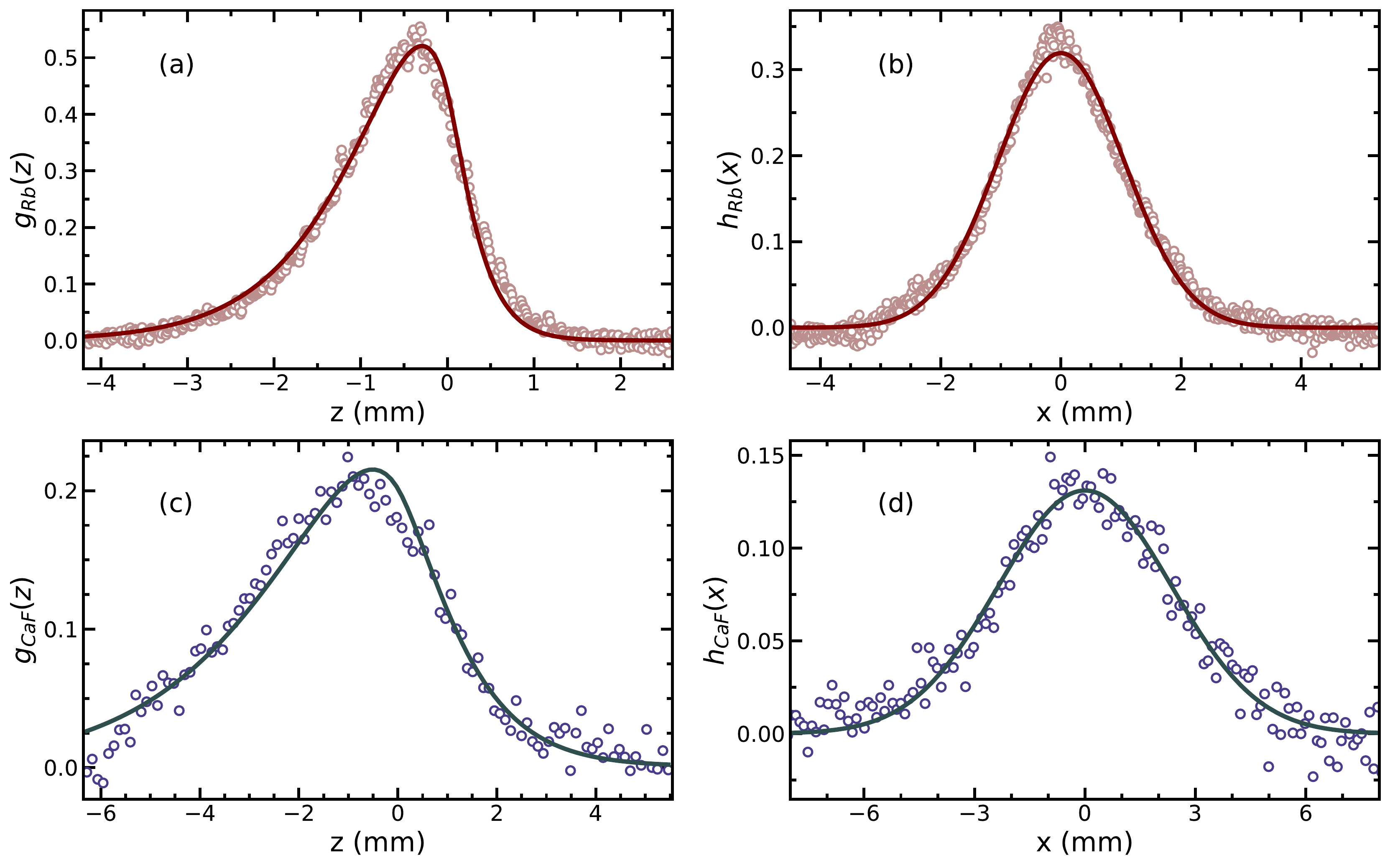}
    \caption{Distributions of Rb and CaF in the magnetic quadrupole trap. (a,c) Axial distributions for (a) Rb and (c) CaF, integrated over the radial direction. Points are data and lines are fits to Eq.~(\ref{eq:1DMQTDist}). (b,d) Radial distributions for (b) Rb and (d) CaF, integrated over the axial direction. Points are data and lines are fits to Eq.~(\ref{eq:gaussDistribution}). The atoms and molecules are held in the magnetic trap for 300~ms, then released and imaged.}
    \label{fig:distributions}
\end{figure*}

In thermal equilibrium at temperature $T$, the density distribution of $N$ atoms of mass $m$ and magnetic moment $\mu$ in a quadrupole magnetic trap with axial field gradient $b$ is
\begin{equation}
    n(x,y,z) = N \frac{\left(\gamma^2 - 1\right)^2}{32\pi z_0^3}\exp\left(-\sqrt{\frac{z^2}{z_0^2}+\frac{x^2+y^2}{4z_0^2}}-\gamma \frac{z}{z_0}\right)
\end{equation}
where $\gamma = m g /\mu b$, $z_0 = k_{\rm B} T/\mu b$, and $g$ is the acceleration due to gravity, which is in the $-z$ direction. Integrating over the $x$ and $y$ dimensions gives the one-dimensional probability density in $z$,
\begin{equation}
    g(z) =  \frac{\left(\gamma^2 - 1\right)^2}{4 z_0}\left(1+\frac{|z|}{z_0}\right)\exp\left(-\frac{|z|+\gamma z}{z_0}\right),
    \label{eq:1DMQTDist}
\end{equation}
normalised such that $\int g(z) dz = 1$.
We fit the axial distributions of both the atoms and molecules to $g(z)$. The data fit reasonably well to this model, and the fits give temperatures close to those measured by the ballistic expansion method, even though the molecules are unlikely to come into thermal equilibrium in the magnetic trap. We find that the radial distributions fit well to Gaussian distributions
\begin{equation}
    h(x) = \frac{1}{\sqrt{2\pi}\sigma}\exp\left(-\frac{x^2}{2\sigma^2}\right).
    \label{eq:gaussDistribution}
\end{equation}
Figure \ref{fig:distributions} shows examples of the axial and radial distributions for both Rb and CaF, together with fits to Eqs.~(\ref{eq:1DMQTDist}) and (\ref{eq:gaussDistribution}).

We model the density distributions of each species as $f_{s}(x,y,z) = h_s(x) h_s(y)g_s(z)$ ($s \in \{\text{CaF},\text{Rb}\}$) using the parameters obtained from the fits. The overlap integral of these distributions is
\begin{align}
    &\int f_{1}(x,y,z)f_{2}(x,y,z)dx\, dy\, dz = \nonumber\\&\frac{A_1^2 A_2^2 \overline{z_0} \left(C^2 \left(4\overline{ z_0}{}^2-3 z_{0,1} z_{0,2}\right)-4 \overline{z_0}{}^2 \left(4\overline{ z_0}{}^2+z_{0,1} z_{0,2}\right)\right)}{4\pi(\sigma_1^2+\sigma_2^2) \left(2 \overline{z_0}+C\right){}^3 \left(B_2 z_{0,1}+B_1 z_{0,2}\right){}^3}, \nonumber
\end{align}
where
\begin{align}
    A_i &= \gamma_i^2 - 1, \nonumber \\
    B_i &= \gamma_i - 1, \nonumber \\
    C &= \gamma_1 z_{0,2}+\gamma_2 z_{0,1}, \nonumber \\ 
    \overline{z_0} &= \frac{1}{2}(z_{0,1}+z_{0,2}). \nonumber
\end{align}
We use this analytical form of the overlap integral to evaluate the effective density $\zeta$ (see Eq.~(\ref{eq:betazeta})). This method accounts for the differential gravitational sag between the two species in a natural way. 

\section{Evaluating uncertainties in loss rates and rate coefficients}

As explained in the main paper, the collision-induced loss rates are determined by fitting the data to Eq.~(\ref{eq:LossModel}) with $r_{\text{CaF}}(0)$ and $\Gamma_{\text{Rb-CaF}}$ as free parameters. The results depend on the values of $\Gamma_{0}^{\text{CaF}}$ and $\Gamma_{\text{Rb}}$, which are measured separately and have their own uncertainties. To account for these uncertainties, we use a re-sampling technique: we draw random values of $r_{\text{CaF}}(t)$ from the set of values measured at each $t$, draw random values of $\Gamma_{\text{Rb}}$ and $\Gamma_{0}^{\text{CaF}}$ from their measured distributions, then fit to Eq.~(\ref{eq:LossModel}). Repeating this several thousand times yields a distribution of $\Gamma_{\text{Rb-CaF}}$ values. These distributions are shown in the insets of Figs.~\ref{fig:loss_N1} and \ref{fig:loss_N0}. The mean and standard deviation of these distributions give the best estimates for $\Gamma_{\text{Rb-CaF}}$.

A similar procedure is used to determine the uncertainties on the measurements of $k_2$. The values of $k_2$ are obtained by fitting a straight line to the graph of $\Gamma_{\text{Rb-CaF}}$ versus $\zeta$, as shown in Fig.~\ref{fig:extract_beta}. We account for the uncertainties in the measured number of atoms and cloud sizes by repeatedly drawing random values of these parameters from their measured distributions and fitting the straight line model each time. This generates a distribution of values of $k_2$, from which we take the mean and standard deviation.

\section{Calculation of loss rates}

Full quantum calculations for systems like Rb+CaF are highly challenging due to the combination of large masses, deep and strongly anisotropic interaction potentials, and electron and nuclear spins. Even if they can be performed, uncertainties in the interaction potentials lead to large uncertainties in cross sections and rates at ultracold temperatures. This happens because the scattering length goes through a full cycle from $-\infty$ to $+\infty$ as an extra bound state is added to or removed from the well.
Therefore, we instead use an approximate single-channel quantum defect theory (QDT) model which accurately reproduces the range of possible loss rates \cite{Frye2015} at much lower computational cost. The model is based on analytic solutions on the asymptotic potential $-C_6R^{-6}$  \cite{Gao1998,Gao:AQDTroutines,Gao2008}, for which we use $C_6=3084\, E_\textrm{h} a_0^6$ as estimated in Ref.\ \cite{Lim2015}. This produces a p-wave barrier height equivalent to 135~$\mu$K.
The complicated short-range physics, including loss processes, is represented simply by an absorbing boundary condition. This is characterized by a non-unitary short-range S-matrix, which can be parameterised as $S=(1-y)(1+y)\exp(\delta^\textrm{s}-\pi/8)$, where $y$ is the loss parameter of Idziaszek and Julienne \cite{Idziaszek2010} and $\delta^\textrm{s}$ is the short-range phase shift. In the absence of loss ($y=0$), this phase shift is related to the scattering length by $a/\bar{a}=1+\cot(\delta^\textrm{s}-\pi/8)$, such that $\delta^\textrm{s}=\pi/8$ corresponds to resonant s-wave scattering. Here, $\bar{a}=0.477988\dots (2\mu C_6/\hbar^2)^{1/4}=35.7$ \AA\ is the mean scattering length of Gribakin and Flambaum \cite{Gribakin1993}.
The asymptotic S-matrices, inelastic scattering cross sections, and loss rate coefficients are calculated as described in Refs.\ \cite{Gao2008, Frye2015}.
Since the two species are at different temperatures, we consider the temperature in the relative motion, which is $T=\mu(T_\textrm{Rb}/m_\textrm{Rb}+T_\textrm{CaF}/m_\textrm{CaF})=133\, \mu$K.
Using this, the loss rates are thermally averaged as $k_2(T)=\int_0^\infty (2/\sqrt{\pi}) k_2(E) x^{1/2} \exp (-x)\,dx$, where $x=E/k_\textrm{B}T$.

To consider whether complex-mediated loss may be a significant factor for this system, we estimate the density of states near threshold. Christianen \emph{et al.} \cite{Christianen(1)2019} give a simple analytic approximate expression for the density of states of diatom+diatom complexes, and we adapt their methodology to apply to atom+diatom systems \cite{FryeUnpublished2021}. We use the estimated parameters for the interaction of Rb and CaF from Ref.\ \cite{Lim2015} and the CaF diatom parameters from \cite{Charron1995}. The resulting density of states is 4~K$^{-1}$, with a corresponding complex lifetime of 60~ps. This is too short to cause significant complex-mediated loss, whether through excitation by a photon or collision with a 3rd body.

%</document>